\documentclass[twocolumn,prb,aps,superscriptaddress,floatfix,citeautoscript,showpacs,10pt]{revtex4-1}
\usepackage{multirow}
\usepackage[dvips]{color,graphicx}
\newcommand{\ca}{H$_{2}$ca}

\begin{document}

\title{Structure and energetics of a ferroelectric organic crystal of
  phenazine and chloranilic acid}
\newcommand{\rutgers}{Department of Physics and Astronomy, Rutgers
  University, Piscataway, New Jersey 08854, USA}
\newcommand{\wfu}{Department of Physics, Wake Forest University,
  Winston-Salem, North Carolina 27109, USA}
\author{Kyuho Lee}\email[Email: ]{klee@physics.rutgers.edu}\affiliation{\rutgers}
\author{Brian Kolb}\affiliation{\wfu}
\author{T. Thonhauser}\affiliation{\wfu}
\author{David Vanderbilt}\affiliation{\rutgers}
\author{David C.~Langreth}\email{Deceased}\affiliation{\rutgers}

\begin{abstract}
  We report first-principles calculations for a ferroelectric organic
  crystal of phenazine and chloranilic acid molecules.  Weak
  intermolecular interactions are properly treated by using a second
  version of van der Waals density functional known as vdW-DF2 [K.~Lee
  \textit{et al}., Phys.~Rev.~B \textbf{82}, 081101 (2010)].  Lattice
  constants, total energies, spontaneous electric polarizations,
  phonon modes and frequencies, and the energy barrier of proton
  transfer are calculated and compared with PBE and experiments
  whenever possible.  We show that the donation of one proton from a
  chloranilic acid molecule to a neighboring phenazine molecule is
  energetically favorable. This proton transfer is the key structural
  change that breaks the centrosymmetry and leads to the ferroelectric
  structure.  However, there is no unstable phonon associated with the
  proton transfer, and an energy barrier of 8\,meV is found between
  the paraelectric and ferroelectric states.
\end{abstract}



\maketitle

\section{Introduction}

Ferroelectric materials are an important class of materials.  Their
responses to various external stimuli can be used for many
applications such as memory devices, electromechanical actuators,
ultrasonic sensors, electro-optic devices, and infrared thermal image
sensors.  Although transition-metal oxides are most widely used for
applications, organic ferroelectrics could be attractive alternatives,
because they could be non-toxic, flexible, and easy to
process.~\cite{Hu2009} However, ferroelectric organic materials are
rare, and substantial efforts are being made to find such materials
that could be of practical use.~\cite{Horiuchi2008} The cocrystal of
phenazine (Phz) and chloranilic acid (\ca) is one of several recently
discovered hydrogen-bonded organic ferroelectrics that have superior
crystallinity and properties compared to conventional ferroelectric
polymers.~\cite{Horiuchi2005, Ishii2006, Asaji2006, Saito2006,
  Gotoh2007, Kumai2007, Horiuchi2008, Horiuchi2009, Fujioka2009,
  Horiuchi2010a, Stroppa2011, Kumai2012}

\begin{figure}[!bp]
  \centering
  \includegraphics[width=0.91\columnwidth]{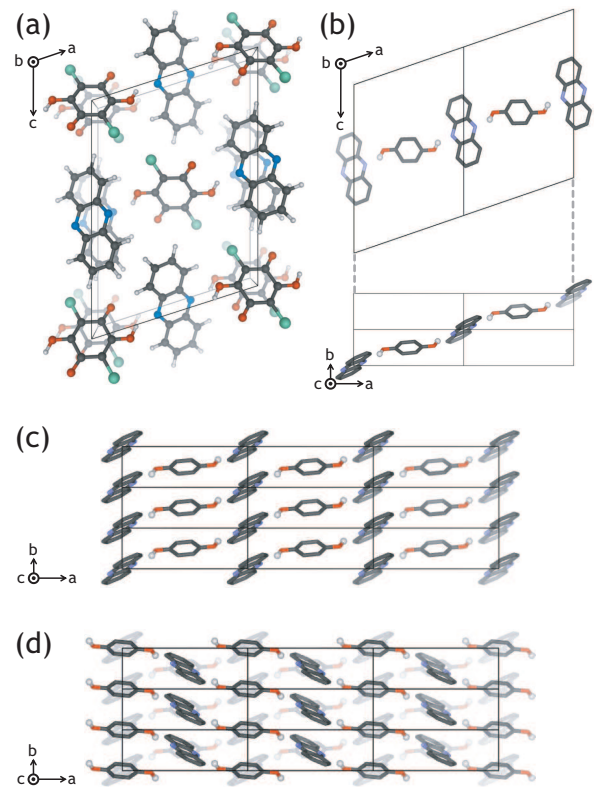}
  \caption{(Color online) (a) Monoclinic unit cell of the paraelectric
    structure containing two \ca\ molecules (at body-center and corner
    sites) and two Phz molecules (at \textit{ab}-face-center and
    $c$-axis edge-center sites).  (b) H-bonded chain of molecules
    running along [110].  (c) An \textit{ab} plane filled by H-bonded
    chains running along [110]. (d) Next higher (or lower) \textit{ab}
    plane filled by H-bonded chains running along $[1\bar{1}0]$.  In
    panels (b-d), all Cl atoms, H atoms bonded to oxygens, and O atoms
    double-bonded to carbons have been omitted for clarity.}
  \label{fig:struc}
\end{figure}

The crystal structure of Phz-\ca\ has been determined by
X-ray~\cite{Horiuchi2005,Gotoh2007} and neutron~\cite{Kumai2007}
diffraction experiments.  The centrosymmetric paraelectric structure
(monoclinic $P2_1/n$, $T>T_\mathrm{c}$ = 253\,K) is shown in
Fig.~\ref{fig:struc}.  There are two molecules of each type (72 atoms
in total) per unit cell. The two constituent molecules, Phz and \ca,
can make hydrogen bonds (H-bonds) with each other, forming linear
chains in the crystal. One such chain that runs along the [110]
direction in an $ab$ plane is shown in Fig.~\ref{fig:struc}(b) as
viewed along the $b$ and $c$ axes (top and bottom subpanels
respectively).  These chains stack along the $b$ direction and fill an
$ab$ plane, as shown in Fig.~\ref{fig:struc}(c).  The next plane above
or below [Fig.~\ref{fig:struc}(d)] is, however, filled with chains
that run along $[1\bar{1}0]$.

Below 253\,K, Phz-\ca\ becomes ferroelectric.~\cite{Horiuchi2005} In
this polar phase (monoclinic $P2_1$), one of the O--H bonds in \ca\
stretches (by about 0.37\,\AA) toward the N atom in the H-bonded Phz
neighbor, adopting the structure shown in
Fig.~\ref{fig:p-order}(b). \cite{Kumai2007} (The large arrows in this
figure represent the directions of the proton displacements.  The
other panels will be discussed later in Sec.~\ref{sec:results}.)  The
polarization, estimated experimentally at 1-2\,$\mu$C/cm$^{2}$, is
parallel to the $b$ axis because the two-fold screw symmetry cancels
out the $a$ and $c$ components of the polarization of each chain.

Interestingly, the proton is found to be almost midway between the
nitrogen and oxygen.  The observed N--H bond length of 1.41\,\AA{} is
much larger than the typical value of 1.03\,\AA{} in proton-transfered
ionic H-bonds.  \cite{Allen1987} It seems that the proton potential
between oxygen and nitrogen has only a single minimum, suggesting that
Phz-\ca\ might not be an order-disorder ferroelectric (FE), but rather
a displacive-type one, i.e., where the paraelectric (PE) phase has a
polar instability associated with a soft phonon mode.  However, a
dynamic proton fluctuation (i.e., a rapid back-and-forth motion of the
proton between two minima located closer to N or O) has been
suggested~\cite{Asaji2006} based on nuclear spin relaxation-time
measurements using the $^{35}$Cl nuclear quadrupole resonance (NQR).
The activation-energy barrier for proton transfer is estimated to be
68\,meV from the Arrhenius temperature dependence of the
fluctuations.~\cite{Asaji2006} Further support for this picture comes
from a second ferroelectric phase (FE-II) that appears upon further
cooling below 136\,K (after passing through an incommensurate phase at
136--146\,K).  In FE-II, the proton is found to be completely
transferred to the Phz nitrogen atom.~\cite{Gotoh2007} The observed
N--H bond length of 1.12\,\AA{} is now consistent with (although a bit
longer than) a typical N--H$^+$ bond length ($\approx$ 1.03\,\AA{}) in
other organic molecular salts.~\cite{Allen1987}

In this work, we investigate the structures, energetics, spontaneous
electric polarizations, lattice instabilities, and energy barriers for
proton transfer by using first-principles calculations.  There are two
difficulties to a theoretical treatment of this important class of
compounds.  First, in order to predict stable crystal structures and
their properties, it is critical to include van der Waals
interactions, which are important for intermolecular interactions but
are missing in conventional exchange-correlation (XC) functionals.
Second, because of the small mass of the proton, the proton quantum
fluctuations are large enough to significantly affect the relative
stabilities of different structures.  To address these issues, we use
the recently developed van der Waals density functional (vdW-DF2) of
Ref.~~\onlinecite{LMKLL10} and include the zero-point energy at the
harmonic level.  The results are compared with experiments as well as
with calculations carried out using the semilocal functional of
Perdew, Burke, and Ernzerhof (PBE)~\cite{PBE96}, one of the most
successful generalized-gradient functionals.  We show that the proton
transfer from \ca\ to Phz lowers the energy by 102\,meV/unit-cell, and
has an energy barrier of  8\,meV. No unstable phonon is associated
with the proton transfer.

The paper is organized as follows.  In Sec.~\ref{sec:methods} we
describe the details of the computational methods used in the
calculations.  Then, in Sec.~\ref{sec:results}, we present the results
of our calculations of lattice constants, energies, and phonons of
polar and nonpolar structures, and of the energy barrier for proton
transfer.  Finally, we summarize the work in Sec.~\ref{sec:summary}.

\section{Methods}
\label{sec:methods}

We use the plane-wave pseudopotential method~\cite{Ihm1979} as
implemented in the \textsc{quantum-espresso} package and
Troullier-Martins norm-conserving pseudopotentials~\cite{TM93}.  We
adopt an kinetic-energy cutoff of 80~Ry and a $2\times6\times1$
$k$-point mesh for the Brillouin-zone sampling.  All calculations are
done fully self-consistently\cite{Thonhauser2007,Langreth2009} using
Soler's efficient algorithm~\cite{RS09} to treat the vdW-DF2
exchange-correlation energy functional.  Atomic positions and lattice
parameters are fully optimized until the residual forces and stresses
are smaller than 7.7\,meV/\AA{} and 0.5 kbar, respectively.  The
Berry-phase technique~\cite{King-Smith1993} is used to calculate the
polarization of the structures.

The transition path for the proton transfer and its energy profile are
determined using the climbing-image nudged elastic band
method~\cite{Henkelman2000,Olsen2004} with 7 images. All atoms are
relaxed during the process, with the lattice parameters fixed to the
experimental equilibrium values for the PE phase.

The zero-point energy corrections are included at the harmonic level
using the computed phonon frequencies.  For the phonon calculation we
use the experimental lattice parameters, the linear-response
density-functional perturbation theory~\cite{Gonze1997,BGD01} for PBE,
and the finite-difference method for vdW-DF2. The acoustic sum rule is
imposed on the force constants.  In order to make the phonon
frequencies well converged up to 1\,cm$^{-1}$, very tight convergence
thresholds are used for the phonon calculation: \texttt{tr2\_ph} and
\texttt{conv\_thr} are set to $10^{-18}$ and $10^{-10}$, respectively,
two orders of magnitude smaller than typical values for inorganic
solids.

The FE and PE states have computed DFT energy gaps of 0.5~eV and
1.2~eV, respectively.  (Since DFT tends to underestimate gaps, the
true gaps are presumably larger.)  All bands are fully occupied and
there are no unpaired electrons.  Thus, Phz-\ca\ is clearly a band
insulator, compatible with our choice of methods.

\section{Results and Discussion}
\label{sec:results}

\subsection{Lattice constants}
\label{sec:latt}

\renewcommand{\arraystretch}{1.2}
\begin{table}[!bp]
  \centering
  \caption{
    Comparison of experimental lattice constants with those
    calculated using PBE and vdW-DF2.
  }
  \label{tab:latt}
  \begin{ruledtabular}
    \begin{tabular}{ccrrr}
      \multirow{2}{*}{Structure} & \multirow{2}{*}{Axis} &
      \multicolumn{3}{c}{Lattice constant (\AA)} \\
      \cline{3-5}
      & & Expt.$^a$ & PBE & vdW-DF2 \\
      \hline 
      \multirow{3}{*}{Paraelectric}
      &      \textit{a} & 12.42 & 11.73 & 12.42 \\
      &      \textit{b} &  3.85 &  4.93 &  3.90 \\
      &      \textit{c} & 16.98 & 16.72 & 16.84 \\
      \hline
      \multirow{3}{*}{Ferroelectric}
      &      \textit{a} & 12.42 & 12.10 & 12.47 \\
      &      \textit{b} &  3.79 &  4.69 &  3.88 \\
      &      \textit{c} & 16.91 & 16.74 & 16.64 \\
    \end{tabular}
  \end{ruledtabular}
  \footnotetext{$^a$Neutron diffraction experiments of
    Ref.~~\onlinecite{Kumai2007}, measured at 300\,K
    and 160\,K for paraelectric and ferroelectric phase respectively.}
\end{table}
\renewcommand{\arraystretch}{1.0}

\begin{figure*}[!tp]
  \centering
  \includegraphics[width=7in]{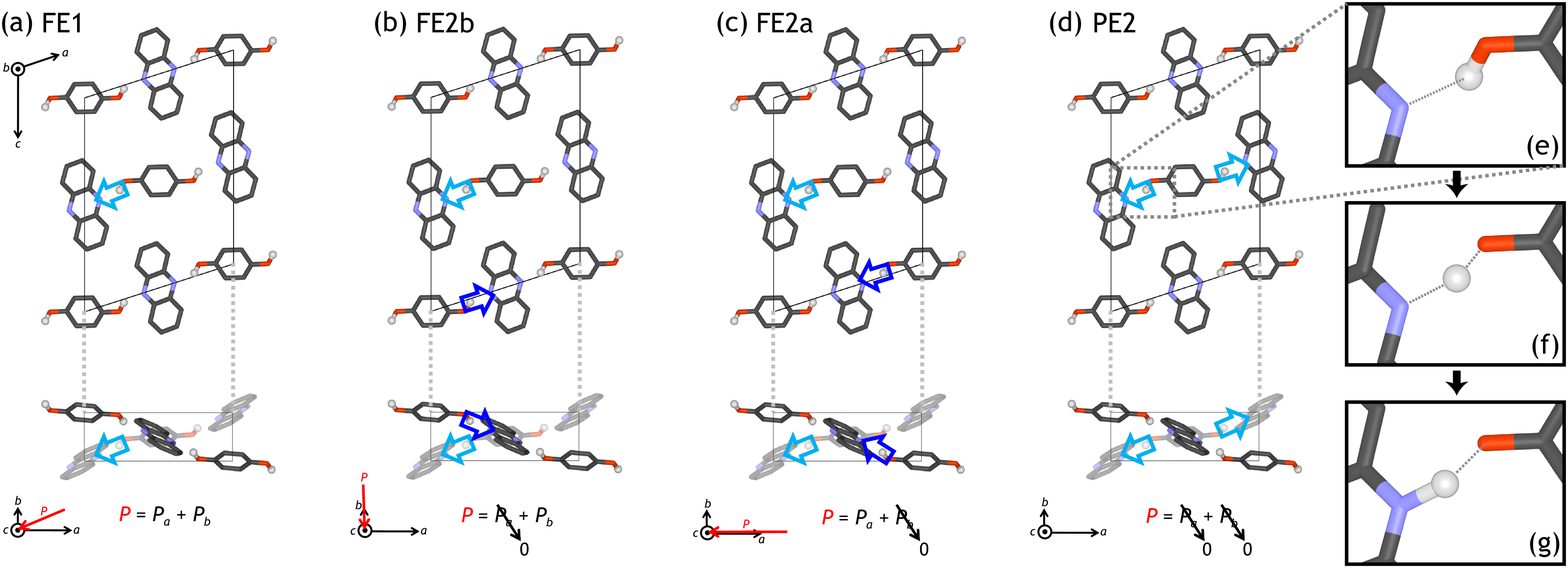}
  \caption{(Color online) Possible proton transfer (PT) processes in
    the unit cell, as indicated by large unfilled arrows.  (a)
    Structure FE1, in which a \ca\ donates a proton to a nitrogen in a
    neighboring Phz, leaving begind an oxygen lone pair on the \ca\
    and creating an electric dipole along the PT direction.  The
    direction of polarization $P$ is shown at the bottom left-hand
    corner as a thin arrow labeled as $P$; here the PT occurs along
    [$\bar{1}\bar{1}$0] so that $P$ has both $a$ and $b$ components.
    (b) Structure FE2b, in which another PT occurs along [1$\bar{1}$0]
    in the other H-bonded chain; $P$ points along $b$ by symmetry.
    (c) Structure FE2a, in which the second PT occurs along
    [$\bar{1}$10] instead; $P$ points along $a$ by symmetry.  (d)
    Structure PE2, a doubly protonated paraelectric structure.
    (e)-(g) Enlarged view of the PT process.}
  \label{fig:p-order}
\end{figure*}

First we validate our computational approach by calculating the
lattice constants and comparing them with known experimental values as
shown in Table~\ref{tab:latt}.  The vdW-DF2 shows excellent agreement
with experiments. For the PE phase, the deviations are 0.00\,\AA{}
(0\%), 0.06\,\AA{} (1\%), and $-0.14$\,\AA{} ($-1$\%) along $a$, $b$,
and $c$, respectively.  The relative deviations with respect to the
experiments are given in the parenthesis.  The deviations for the FE
phase are similar: 0.05\,\AA{} (0\%), 0.09\,\AA{} (2\%), and
$-0.27$\,\AA{} ($-2$\%).

On the other hand, PBE, which is one of the most successful semi-local
functionals, overestimates the PE lattice constants by $-0.70$\,\AA{}
($-6$\%), 1.08\,\AA{} (28\%) and $-0.26$\,\AA{} ($-2$\%) along $a$,
$b$, and $c$ axis, respectively [for FE, $-0.33$\,\AA{} ($-3$\%),
0.90\,\AA{} (24\%) and $-0.17$\,\AA{} ($-1$\%)].  Except for the $c$
lattice constant for the PE phase, all lattice constants are poorly
reproduced.

This comparison between vdW-DF2 predictions and experiments confirms
that the vdW-DF2 functional is capable of capturing all three
important interactions (covalent bonds, H-bonds, and van der Waals
interactions) with good fidelity in this organic crystal.
%

\subsection{Proton-transferred structures}

In the PE phase, there is no proton transfer and all molecules are
neutral.  There are two \ca\ molecules per unit cell, and each \ca\
has two hydrogen bonds.  The protons in those four hydrogen bonds can
be transferred to neighboring Phz molecules.  In this section, we
consider all the possible proton transfer configurations consistent
with the primitive-cell periodicity, starting from the simplest and
working toward more complex ones.

The simplest single-proton transfer is shown in
Fig.~\ref{fig:p-order}(a), where the \ca\ at the center of the unit
cell donates a proton to a neighboring Phz (to the left in this
figure).  The large arrow represents the direction of the proton
transfer.  We denote the resulting structure as FE1, indicating that
it is ferroelectric and only one proton has transferred in the unit
cell.

\begin{figure}[!tp]
  \centering
  \includegraphics[width=.68\columnwidth]{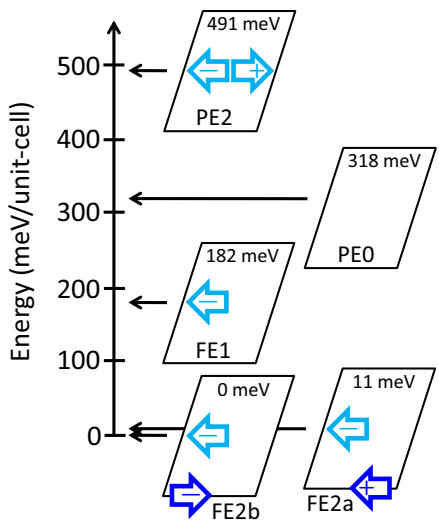}
  \caption{(Color online) Energies of competitive proton-transferred
    structures presented relative to the FE2b ground-state energy,
    which is taken as zero.}
  \label{fig:energy}
\end{figure}

This structural change breaks the centrosymmetry, making it
ferroelectric, and lowers the energy by 137\,meV per unit cell with
respect to the paraelectric phase, which we denote henceforth as PE0.
The relative energies of these and other structures (to be discussed
shortly) are illustrated in Fig.~\ref{fig:energy}.  (All structures
and energies in this section are calculated using vdW-DF2.  A
comparison with PBE will be given later.)  The electric polarization
$P$ of the FE1 structure points approximately in the [110] direction
as is shown by the thin arrow at the bottom left-hand corner of the
figure.

The [H$_1$ca]$^-$ molecule that already donated one proton also has
the possibility to donate a second one to the other neighboring Phz in
the chain, as shown in Fig.~\ref{fig:p-order}(d).  We denote this
structure as PE2, where the `2' indicates that two protons are
transferred in the unit cell.  This double protonation of a Phz
restores the centrosymmetry, so that PE2 has no polarization.  However
this second proton transfer increases the energy by 309\,meV, i.e.,
the energy of PE2 is 172\,meV higher than that of PE.  Therefore, in
searching for the ground state, we do not consider other
configurations based on double-protonation chains, i.e., those
involving three (FE3) or four (FE4) transferred protons per cell.

Two possibilities then remain, in which the other \ca\ molecule at the
corner site in the figure also donates a proton, as shown in
Figs.~\ref{fig:p-order}(b-c).  We found that the experimental
low-temperature ferroelectric structure that we already discussed in
detail in the previous section, shown in Fig.~\ref{fig:p-order}(b),
has the lowest energy of all possible configurations.  We now denote
it as FE2$b$, indicating that two protons are transferred and the net
polarization is parallel to the `$b$' axis.  FE2b is more stable than
PE0 by 318\,meV.  This energy reduction is 44\,meV larger than twice
the PE0-to-FE1 reduction of 137\,meV mentioned above.  Thus, 44\,meV
can be taken as an estimate of the interchain coupling strength along
$c$.

The polarization of FE2b is calculated to be
4.5\,$\mathrm{\mu{}C/cm^{2}}$, in a reasonable agreement with the
measured value of $\sim$2\,$\mathrm{\mu{}C/cm^{2}}$.  Also our
calculated N--H bond length of 1.06\,\AA{} is consistent with the
experimental value of 1.12\,\AA{} measured at low temperature in phase
FE-II, and with a typical N--H$^+$ bond length of $\approx$1.03\,\AA{}
in other organic molecular salts.~\cite{Allen1987} Thus, the structure
of the ferroelectric phase agrees also well with experiments.

In the other case, in which the proton is transferred to the left Phz
as in Fig.~\ref{fig:p-order}(c), only the $a$ component of the
polarization remains by the symmetry.  Following the same notational
scheme introduced above, we denote this as FE2$a$.  Its energy is only
11\,meV/unit-cell higher than that of the ground-state FE2$b$
structure.

\begin{figure}[t]
  \centering
  \includegraphics[width=1.0\columnwidth]{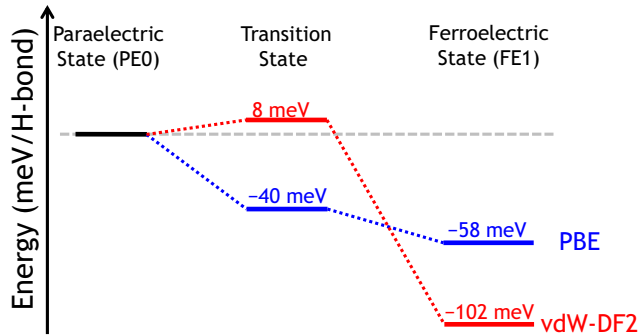}
  \caption{ (Color online) Energies relative to PE0 (left), as
    calculated with PBE and vdW-DF2, for the transition-state
    structure leading from PE0 to FE1 (middle) and for FE1 (right).
    The lattice constants are fixed to experimental values for this
    comparison, and zero-point energies are included at the harmonic
    level.  }
  \label{fig:barrier}
\end{figure}

\subsection{Energy barrier for proton transfer and lattice
  instability}

Here we investigate the proton transfer process from PE0 to FE1 in
detail.  The transition path for the proton transfer is calculated
using the climbing-image nudged elastic band
method.~\cite{Henkelman2000,Olsen2004} A close-up view of the initial,
transition, and final states of this proton transfer process are shown
in Figs.~\ref{fig:p-order}(e)-(g).  Along the transition path, the two
molecules get closer to each other by up to 0.18\,\AA. They then
retreat again as the proton completes its transfer, but not
completely; the final proton-transferred pair is closer by 0.05\,\AA{}
than the initial neutral one.

It is well known that local or semi-local functionals such as PBE
underestimate proton-transfer barriers.~\cite{Sadhukhan1999,Zhao2004}
For example, for the intramolecular proton transfer in the
malonaldehyde molecule, the energy barrier in PBE is only a quater of
the known accurate value of 177\,meV
(Ref.~~\onlinecite{Sadhukhan1999}) calculated by the coupled-cluster
method with single, double, and perturbative triple exciations
[CCSD(T)].  On the other hand, the vdW-DF2 barrier of 183\,meV in that
case is in excellent agreement with the accurate value, corroborating
the validity of our method.  For the Phz-\ca\ crystal, we found a
similar result; our computed PBE barrier of 44\,meV from PE0 to FE1 is
less than half of the vdW-DF2 value (105\,meV).  Furthermore, after
including the zero-point corrections, the vdW-DF2 barrier drops to
8\,meV, and the transition becomes barrierless in PBE as is shown in
Fig.~\ref{fig:barrier}.

The energetics of proton transfer in vdW-DF2 is in good agreement with
experimental observation of the thermally activated proton fluctuation
in the high-temperature FE-I phase and the proton-transferred
structure in the low-temperature FE-II phase.  Also the apparent
single-well proton potential in FE-I could be understood by dynamic
proton transfer in a small-barrier double-well potential.

Nevertheless, there could be other pathways not captured by our
initial NEB path, or an unstable phonon mode that could trigger the
proton displacement toward the FE-I phase.  However, our calculation
shows that the zone-center phonons of the PE0 structure are all
stable.  We do not find any unstable mode associated with the proton
displacement toward the FE-I phase, the signature of a displacive-type
ferroelectric.

Both the proton transfer energetics and the zone-center phonons are in
good agreement with experimental observations and support the picture
of an order-disorder, as opposed to a displacive, FE transition.

\section{Conclusion}
\label{sec:summary}

By using first-principles density-functional theory, we have studied
the structure and energetics of a ferroelectric molecular crystal of
phenazine and chloranilic acid, and have analyzed the energy barrier
for proton transfer and the stability of lattice vibrational modes.
We have shown that the inclusion of van der Waals interactions is
crucial for a proper description of this molecular crystal, and that a
recently developed vdW-DF2 functional reproduces the structures of the
PE and FE phases in good agreement with experiment.  We have found
that the zone-center phonons of the PE state are all stable, and the
proton transfer---the key structural change that leads to the
ferroelectric structure---has an energy barrier of 8\,meV.  The
signature of a displacive-type ferroelectric, i.e., lattice
instability in the PE phase, has not been found.  Accordingly we
propose that Phz-\ca\ is an order-disorder FE.  Our analysis of the
stability of the lattice vibrational modes and the energy barrier for
proton transfer supports the possibility that the apparent single-well
proton potential in the FE-I phase is an average effect arising from
dynamical proton transfer in an asymmetric double-well potential.

\section*{Acknowledgments}

We thank K.~Rabe, J.~H.~Lee, A.~Kumar, and S.~Coh for useful
discussions.  The work at Rutgers supported by the National Science
Foundation under Grant number DMR-0801343.

\bibliographystyle{apsrev4-1}
\bibliography{ofe}

\begin{thebibliography}{28}%
\makeatletter
\providecommand \@ifxundefined [1]{%
 \@ifx{#1\undefined}
}%
\providecommand \@ifnum [1]{%
 \ifnum #1\expandafter \@firstoftwo
 \else \expandafter \@secondoftwo
 \fi
}%
\providecommand \@ifx [1]{%
 \ifx #1\expandafter \@firstoftwo
 \else \expandafter \@secondoftwo
 \fi
}%
\providecommand \natexlab [1]{#1}%
\providecommand \enquote  [1]{``#1''}%
\providecommand \bibnamefont  [1]{#1}%
\providecommand \bibfnamefont [1]{#1}%
\providecommand \citenamefont [1]{#1}%
\providecommand \href@noop [0]{\@secondoftwo}%
\providecommand \href [0]{\begingroup \@sanitize@url \@href}%
\providecommand \@href[1]{\@@startlink{#1}\@@href}%
\providecommand \@@href[1]{\endgroup#1\@@endlink}%
\providecommand \@sanitize@url [0]{\catcode `\\12\catcode `\$12\catcode
  `\&12\catcode `\#12\catcode `\^12\catcode `\_12\catcode `\%12\relax}%
\providecommand \@@startlink[1]{}%
\providecommand \@@endlink[0]{}%
\providecommand \url  [0]{\begingroup\@sanitize@url \@url }%
\providecommand \@url [1]{\endgroup\@href {#1}{\urlprefix }}%
\providecommand \urlprefix  [0]{URL }%
\providecommand \Eprint [0]{\href }%
\providecommand \doibase [0]{http://dx.doi.org/}%
\providecommand \selectlanguage [0]{\@gobble}%
\providecommand \bibinfo  [0]{\@secondoftwo}%
\providecommand \bibfield  [0]{\@secondoftwo}%
\providecommand \translation [1]{[#1]}%
\providecommand \BibitemOpen [0]{}%
\providecommand \bibitemStop [0]{}%
\providecommand \bibitemNoStop [0]{.\EOS\space}%
\providecommand \EOS [0]{\spacefactor3000\relax}%
\providecommand \BibitemShut  [1]{\csname bibitem#1\endcsname}%
\let\auto@bib@innerbib\@empty
\bibitem [{\citenamefont {Hu}\ \emph {et~al.}(2009)\citenamefont {Hu},
  \citenamefont {Tian}, \citenamefont {Nysten},\ and\ \citenamefont
  {Jonas}}]{Hu2009}%
  \BibitemOpen
  \bibfield  {author} {\bibinfo {author} {\bibfnamefont {Z.}~\bibnamefont
  {Hu}}, \bibinfo {author} {\bibfnamefont {M.}~\bibnamefont {Tian}}, \bibinfo
  {author} {\bibfnamefont {B.}~\bibnamefont {Nysten}}, \ and\ \bibinfo {author}
  {\bibfnamefont {A.~M.}\ \bibnamefont {Jonas}},\ }\href {\doibase
  10.1038/nmat2339} {\bibfield  {journal} {\bibinfo  {journal} {Nat. Mater.}\
  }\textbf {\bibinfo {volume} {8}},\ \bibinfo {pages} {62} (\bibinfo {year}
  {2009})}\BibitemShut {NoStop}%
\bibitem [{\citenamefont {Horiuchi}\ and\ \citenamefont
  {Tokura}(2008)}]{Horiuchi2008}%
  \BibitemOpen
  \bibfield  {author} {\bibinfo {author} {\bibfnamefont {S.}~\bibnamefont
  {Horiuchi}}\ and\ \bibinfo {author} {\bibfnamefont {Y.}~\bibnamefont
  {Tokura}},\ }\href {\doibase 10.1038/nmat2137} {\bibfield  {journal}
  {\bibinfo  {journal} {Nat. Mater.}\ }\textbf {\bibinfo {volume} {7}},\
  \bibinfo {pages} {357} (\bibinfo {year} {2008})}\BibitemShut {NoStop}%
\bibitem [{\citenamefont {Horiuchi}\ \emph {et~al.}(2005)\citenamefont
  {Horiuchi}, \citenamefont {Ishii}, \citenamefont {Kumai}, \citenamefont
  {Okimoto}, \citenamefont {Tachibana}, \citenamefont {Nagaosa},\ and\
  \citenamefont {Tokura}}]{Horiuchi2005}%
  \BibitemOpen
  \bibfield  {author} {\bibinfo {author} {\bibfnamefont {S.}~\bibnamefont
  {Horiuchi}}, \bibinfo {author} {\bibfnamefont {F.}~\bibnamefont {Ishii}},
  \bibinfo {author} {\bibfnamefont {R.}~\bibnamefont {Kumai}}, \bibinfo
  {author} {\bibfnamefont {Y.}~\bibnamefont {Okimoto}}, \bibinfo {author}
  {\bibfnamefont {H.}~\bibnamefont {Tachibana}}, \bibinfo {author}
  {\bibfnamefont {N.}~\bibnamefont {Nagaosa}}, \ and\ \bibinfo {author}
  {\bibfnamefont {Y.}~\bibnamefont {Tokura}},\ }\href {\doibase
  10.1038/nmat1298} {\bibfield  {journal} {\bibinfo  {journal} {Nat. Mater.}\
  }\textbf {\bibinfo {volume} {4}},\ \bibinfo {pages} {163} (\bibinfo {year}
  {2005})}\BibitemShut {NoStop}%
\bibitem [{\citenamefont {Ishii}\ \emph {et~al.}(2006)\citenamefont {Ishii},
  \citenamefont {Nagaosa}, \citenamefont {Tokura},\ and\ \citenamefont
  {Terakura}}]{Ishii2006}%
  \BibitemOpen
  \bibfield  {author} {\bibinfo {author} {\bibfnamefont {F.}~\bibnamefont
  {Ishii}}, \bibinfo {author} {\bibfnamefont {N.}~\bibnamefont {Nagaosa}},
  \bibinfo {author} {\bibfnamefont {Y.}~\bibnamefont {Tokura}}, \ and\ \bibinfo
  {author} {\bibfnamefont {K.}~\bibnamefont {Terakura}},\ }\href@noop {}
  {\bibfield  {journal} {\bibinfo  {journal} {Phys. Rev. B}\ }\textbf {\bibinfo
  {volume} {73}},\ \bibinfo {pages} {212105} (\bibinfo {year}
  {2006})}\BibitemShut {NoStop}%
\bibitem [{\citenamefont {Asaji}\ \emph {et~al.}(2006)\citenamefont {Asaji},
  \citenamefont {Gotoh},\ and\ \citenamefont {Watanabe}}]{Asaji2006}%
  \BibitemOpen
  \bibfield  {author} {\bibinfo {author} {\bibfnamefont {T.}~\bibnamefont
  {Asaji}}, \bibinfo {author} {\bibfnamefont {K.}~\bibnamefont {Gotoh}}, \ and\
  \bibinfo {author} {\bibfnamefont {J.}~\bibnamefont {Watanabe}},\ }\href
  {\doibase 10.1016/j.molstruc.2006.01.006} {\bibfield  {journal} {\bibinfo
  {journal} {J. Mol. Struct.}\ }\textbf {\bibinfo {volume} {791}},\ \bibinfo
  {pages} {89} (\bibinfo {year} {2006})}\BibitemShut {NoStop}%
\bibitem [{\citenamefont {Saito}\ \emph {et~al.}(2006)\citenamefont {Saito},
  \citenamefont {Amano}, \citenamefont {Yamamura}, \citenamefont {Tojo},\ and\
  \citenamefont {Atake}}]{Saito2006}%
  \BibitemOpen
  \bibfield  {author} {\bibinfo {author} {\bibfnamefont {K.}~\bibnamefont
  {Saito}}, \bibinfo {author} {\bibfnamefont {M.}~\bibnamefont {Amano}},
  \bibinfo {author} {\bibfnamefont {Y.}~\bibnamefont {Yamamura}}, \bibinfo
  {author} {\bibfnamefont {T.}~\bibnamefont {Tojo}}, \ and\ \bibinfo {author}
  {\bibfnamefont {T.}~\bibnamefont {Atake}},\ }\href {\doibase
  10.1143/JPSJ.75.033601} {\bibfield  {journal} {\bibinfo  {journal} {J. Phys.
  Soc. Jpn.}\ }\textbf {\bibinfo {volume} {75}},\ \bibinfo {pages} {033601}
  (\bibinfo {year} {2006})}\BibitemShut {NoStop}%
\bibitem [{\citenamefont {Gotoh}\ \emph {et~al.}(2007)\citenamefont {Gotoh},
  \citenamefont {Asaji},\ and\ \citenamefont {Ishida}}]{Gotoh2007}%
  \BibitemOpen
  \bibfield  {author} {\bibinfo {author} {\bibfnamefont {K.}~\bibnamefont
  {Gotoh}}, \bibinfo {author} {\bibfnamefont {T.}~\bibnamefont {Asaji}}, \ and\
  \bibinfo {author} {\bibfnamefont {H.}~\bibnamefont {Ishida}},\ }\href
  {\doibase 10.1107/S0108270106049468} {\bibfield  {journal} {\bibinfo
  {journal} {Acta Crystallogr. C}\ }\textbf {\bibinfo {volume} {63}},\ \bibinfo
  {pages} {o17} (\bibinfo {year} {2007})}\BibitemShut {NoStop}%
\bibitem [{\citenamefont {Kumai}\ \emph {et~al.}(2007)\citenamefont {Kumai},
  \citenamefont {Horiuchi}, \citenamefont {Sagayama}, \citenamefont {Arima},
  \citenamefont {Watanabe}, \citenamefont {Noda},\ and\ \citenamefont
  {Tokura}}]{Kumai2007}%
  \BibitemOpen
  \bibfield  {author} {\bibinfo {author} {\bibfnamefont {R.}~\bibnamefont
  {Kumai}}, \bibinfo {author} {\bibfnamefont {S.}~\bibnamefont {Horiuchi}},
  \bibinfo {author} {\bibfnamefont {H.}~\bibnamefont {Sagayama}}, \bibinfo
  {author} {\bibfnamefont {T.-H.}\ \bibnamefont {Arima}}, \bibinfo {author}
  {\bibfnamefont {M.}~\bibnamefont {Watanabe}}, \bibinfo {author}
  {\bibfnamefont {Y.}~\bibnamefont {Noda}}, \ and\ \bibinfo {author}
  {\bibfnamefont {Y.}~\bibnamefont {Tokura}},\ }\href {\doibase
  10.1021/ja075406r} {\bibfield  {journal} {\bibinfo  {journal} {J. Am. Chem.
  Soc.}\ }\textbf {\bibinfo {volume} {129}},\ \bibinfo {pages} {12920}
  (\bibinfo {year} {2007})}\BibitemShut {NoStop}%
\bibitem [{\citenamefont {Horiuchi}\ \emph {et~al.}(2009)\citenamefont
  {Horiuchi}, \citenamefont {Kumai},\ and\ \citenamefont
  {Tokura}}]{Horiuchi2009}%
  \BibitemOpen
  \bibfield  {author} {\bibinfo {author} {\bibfnamefont {S.}~\bibnamefont
  {Horiuchi}}, \bibinfo {author} {\bibfnamefont {R.}~\bibnamefont {Kumai}}, \
  and\ \bibinfo {author} {\bibfnamefont {Y.}~\bibnamefont {Tokura}},\ }\href
  {\doibase 10.1039/b900987f} {\bibfield  {journal} {\bibinfo  {journal} {J.
  Mater. Chem.}\ }\textbf {\bibinfo {volume} {19}},\ \bibinfo {pages} {4421}
  (\bibinfo {year} {2009})}\BibitemShut {NoStop}%
\bibitem [{\citenamefont {Fujioka}\ \emph {et~al.}(2009)\citenamefont
  {Fujioka}, \citenamefont {Horiuchi}, \citenamefont {Kida}, \citenamefont
  {Shimano},\ and\ \citenamefont {Tokura}}]{Fujioka2009}%
  \BibitemOpen
  \bibfield  {author} {\bibinfo {author} {\bibfnamefont {J.}~\bibnamefont
  {Fujioka}}, \bibinfo {author} {\bibfnamefont {S.}~\bibnamefont {Horiuchi}},
  \bibinfo {author} {\bibfnamefont {N.}~\bibnamefont {Kida}}, \bibinfo {author}
  {\bibfnamefont {R.}~\bibnamefont {Shimano}}, \ and\ \bibinfo {author}
  {\bibfnamefont {Y.}~\bibnamefont {Tokura}},\ }\href {\doibase
  10.1103/PhysRevB.80.125134} {\bibfield  {journal} {\bibinfo  {journal} {Phys.
  Rev. B}\ }\textbf {\bibinfo {volume} {80}},\ \bibinfo {pages} {125134}
  (\bibinfo {year} {2009})}\BibitemShut {NoStop}%
\bibitem [{\citenamefont {Horiuchi}\ \emph {et~al.}(2010)\citenamefont
  {Horiuchi}, \citenamefont {Kumai}, \citenamefont {Fujioka},\ and\
  \citenamefont {Tokura}}]{Horiuchi2010a}%
  \BibitemOpen
  \bibfield  {author} {\bibinfo {author} {\bibfnamefont {S.}~\bibnamefont
  {Horiuchi}}, \bibinfo {author} {\bibfnamefont {R.}~\bibnamefont {Kumai}},
  \bibinfo {author} {\bibfnamefont {J.}~\bibnamefont {Fujioka}}, \ and\
  \bibinfo {author} {\bibfnamefont {Y.}~\bibnamefont {Tokura}},\ }\href
  {\doibase 10.1016/j.physb.2009.10.021} {\bibfield  {journal} {\bibinfo
  {journal} {Physica B}\ }\textbf {\bibinfo {volume} {405}},\ \bibinfo {pages}
  {S334} (\bibinfo {year} {2010})}\BibitemShut {NoStop}%
\bibitem [{\citenamefont {Stroppa}\ \emph {et~al.}(2011)\citenamefont
  {Stroppa}, \citenamefont {Di~Sante}, \citenamefont {Horiuchi}, \citenamefont
  {Tokura}, \citenamefont {Vanderbilt},\ and\ \citenamefont
  {Picozzi}}]{Stroppa2011}%
  \BibitemOpen
  \bibfield  {author} {\bibinfo {author} {\bibfnamefont {A.}~\bibnamefont
  {Stroppa}}, \bibinfo {author} {\bibfnamefont {D.}~\bibnamefont {Di~Sante}},
  \bibinfo {author} {\bibfnamefont {S.}~\bibnamefont {Horiuchi}}, \bibinfo
  {author} {\bibfnamefont {Y.}~\bibnamefont {Tokura}}, \bibinfo {author}
  {\bibfnamefont {D.}~\bibnamefont {Vanderbilt}}, \ and\ \bibinfo {author}
  {\bibfnamefont {S.}~\bibnamefont {Picozzi}},\ }\href@noop {} {\bibfield
  {journal} {\bibinfo  {journal} {Phys. Rev. B}\ }\textbf {\bibinfo {volume}
  {84}},\ \bibinfo {pages} {014101} (\bibinfo {year} {2011})}\BibitemShut
  {NoStop}%
\bibitem [{\citenamefont {Kumai}\ \emph {et~al.}(2012)\citenamefont {Kumai},
  \citenamefont {Horiuchi}, \citenamefont {Fujioka},\ and\ \citenamefont
  {Tokura}}]{Kumai2012}%
  \BibitemOpen
  \bibfield  {author} {\bibinfo {author} {\bibfnamefont {R.}~\bibnamefont
  {Kumai}}, \bibinfo {author} {\bibfnamefont {S.}~\bibnamefont {Horiuchi}},
  \bibinfo {author} {\bibfnamefont {J.}~\bibnamefont {Fujioka}}, \ and\
  \bibinfo {author} {\bibfnamefont {Y.}~\bibnamefont {Tokura}},\ }\href
  {\doibase 10.1021/ja208113p} {\bibfield  {journal} {\bibinfo  {journal} {J.
  Am. Chem. Soc.}\ }\textbf {\bibinfo {volume} {134}},\ \bibinfo {pages} {1036}
  (\bibinfo {year} {2012})}\BibitemShut {NoStop}%
\bibitem [{\citenamefont {Allen}\ \emph {et~al.}(1987)\citenamefont {Allen},
  \citenamefont {Kennard}, \citenamefont {Watson}, \citenamefont {Brammer},
  \citenamefont {Orpen},\ and\ \citenamefont {Taylor}}]{Allen1987}%
  \BibitemOpen
  \bibfield  {author} {\bibinfo {author} {\bibfnamefont {F.~H.}\ \bibnamefont
  {Allen}}, \bibinfo {author} {\bibfnamefont {O.}~\bibnamefont {Kennard}},
  \bibinfo {author} {\bibfnamefont {D.~G.}\ \bibnamefont {Watson}}, \bibinfo
  {author} {\bibfnamefont {L.}~\bibnamefont {Brammer}}, \bibinfo {author}
  {\bibfnamefont {A.~G.}\ \bibnamefont {Orpen}}, \ and\ \bibinfo {author}
  {\bibfnamefont {R.}~\bibnamefont {Taylor}},\ }\href {\doibase
  10.1039/p298700000s1} {\bibfield  {journal} {\bibinfo  {journal} {J. Chem.
  Soc. Perkin Trans. II}\ ,\ \bibinfo {pages} {S1}} (\bibinfo {year}
  {1987})}\BibitemShut {NoStop}%
\bibitem [{\citenamefont {Lee}\ \emph {et~al.}(2010)\citenamefont {Lee},
  \citenamefont {Murray}, \citenamefont {Kong}, \citenamefont {Lundqvist},\
  and\ \citenamefont {Langreth}}]{LMKLL10}%
  \BibitemOpen
  \bibfield  {author} {\bibinfo {author} {\bibfnamefont {K.}~\bibnamefont
  {Lee}}, \bibinfo {author} {\bibfnamefont {E.~D.}\ \bibnamefont {Murray}},
  \bibinfo {author} {\bibfnamefont {L.}~\bibnamefont {Kong}}, \bibinfo {author}
  {\bibfnamefont {B.~I.}\ \bibnamefont {Lundqvist}}, \ and\ \bibinfo {author}
  {\bibfnamefont {D.~C.}\ \bibnamefont {Langreth}},\ }\href {\doibase
  10.1103/PhysRevB.82.081101} {\bibfield  {journal} {\bibinfo  {journal} {Phys.
  Rev. B}\ }\textbf {\bibinfo {volume} {82}},\ \bibinfo {pages} {081101}
  (\bibinfo {year} {2010})}\BibitemShut {NoStop}%
\bibitem [{\citenamefont {Perdew}\ \emph {et~al.}(1996)\citenamefont {Perdew},
  \citenamefont {Burke},\ and\ \citenamefont {Ernzerhof}}]{PBE96}%
  \BibitemOpen
  \bibfield  {author} {\bibinfo {author} {\bibfnamefont {J.~P.}\ \bibnamefont
  {Perdew}}, \bibinfo {author} {\bibfnamefont {K.}~\bibnamefont {Burke}}, \
  and\ \bibinfo {author} {\bibfnamefont {M.}~\bibnamefont {Ernzerhof}},\ }\href
  {\doibase 10.1103/PhysRevLett.77.3865} {\bibfield  {journal} {\bibinfo
  {journal} {Phys. Rev. Lett.}\ }\textbf {\bibinfo {volume} {77}},\ \bibinfo
  {pages} {3865} (\bibinfo {year} {1996})}\BibitemShut {NoStop}%
\bibitem [{\citenamefont {Ihm}\ \emph {et~al.}(1979)\citenamefont {Ihm},
  \citenamefont {Zunger},\ and\ \citenamefont {Cohen}}]{Ihm1979}%
  \BibitemOpen
  \bibfield  {author} {\bibinfo {author} {\bibfnamefont {J.}~\bibnamefont
  {Ihm}}, \bibinfo {author} {\bibfnamefont {A.}~\bibnamefont {Zunger}}, \ and\
  \bibinfo {author} {\bibfnamefont {M.~L.}\ \bibnamefont {Cohen}},\ }\href
  {\doibase 10.1088/0022-3719/12/21/009} {\bibfield  {journal} {\bibinfo
  {journal} {J. Phys. C: Solid State Phys.}\ }\textbf {\bibinfo {volume}
  {12}},\ \bibinfo {pages} {4409} (\bibinfo {year} {1979})}\BibitemShut
  {NoStop}%
\bibitem [{\citenamefont {Troullier}\ and\ \citenamefont
  {Martins}(1991)}]{TM93}%
  \BibitemOpen
  \bibfield  {author} {\bibinfo {author} {\bibfnamefont {N.}~\bibnamefont
  {Troullier}}\ and\ \bibinfo {author} {\bibfnamefont {J.~L.}\ \bibnamefont
  {Martins}},\ }\href {\doibase 10.1103/PhysRevB.43.1993} {\bibfield  {journal}
  {\bibinfo  {journal} {Phys. Rev. B}\ }\textbf {\bibinfo {volume} {43}},\
  \bibinfo {pages} {1993} (\bibinfo {year} {1991})}\BibitemShut {NoStop}%
\bibitem [{\citenamefont {Thonhauser}\ \emph {et~al.}(2007)\citenamefont
  {Thonhauser}, \citenamefont {Cooper}, \citenamefont {Li}, \citenamefont
  {Puzder}, \citenamefont {Hyldgaard},\ and\ \citenamefont
  {Langreth}}]{Thonhauser2007}%
  \BibitemOpen
  \bibfield  {author} {\bibinfo {author} {\bibfnamefont {T.}~\bibnamefont
  {Thonhauser}}, \bibinfo {author} {\bibfnamefont {V.~R.}\ \bibnamefont
  {Cooper}}, \bibinfo {author} {\bibfnamefont {S.}~\bibnamefont {Li}}, \bibinfo
  {author} {\bibfnamefont {A.}~\bibnamefont {Puzder}}, \bibinfo {author}
  {\bibfnamefont {P.}~\bibnamefont {Hyldgaard}}, \ and\ \bibinfo {author}
  {\bibfnamefont {D.~C.}\ \bibnamefont {Langreth}},\ }\href@noop {} {\bibfield
  {journal} {\bibinfo  {journal} {Phys. Rev. B}\ }\textbf {\bibinfo {volume}
  {76}},\ \bibinfo {pages} {125112} (\bibinfo {year} {2007})}\BibitemShut
  {NoStop}%
\bibitem [{\citenamefont {Langreth}\ \emph {et~al.}(2009)\citenamefont
  {Langreth}, \citenamefont {Lundqvist}, \citenamefont {Chakarova-Käck},
  \citenamefont {Cooper}, \citenamefont {Dion}, \citenamefont {Hyldgaard},
  \citenamefont {Kelkkanen}, \citenamefont {Kleis}, \citenamefont {Kong},
  \citenamefont {Li}, \citenamefont {Moses}, \citenamefont {Murray},
  \citenamefont {Puzder}, \citenamefont {Rydberg}, \citenamefont {Schröder},\
  and\ \citenamefont {Thonhauser}}]{Langreth2009}%
  \BibitemOpen
  \bibfield  {author} {\bibinfo {author} {\bibfnamefont {D.~C.}\ \bibnamefont
  {Langreth}}, \bibinfo {author} {\bibfnamefont {B.~I.}\ \bibnamefont
  {Lundqvist}}, \bibinfo {author} {\bibfnamefont {S.~D.}\ \bibnamefont
  {Chakarova-Käck}}, \bibinfo {author} {\bibfnamefont {V.~R.}\ \bibnamefont
  {Cooper}}, \bibinfo {author} {\bibfnamefont {M.}~\bibnamefont {Dion}},
  \bibinfo {author} {\bibfnamefont {P.}~\bibnamefont {Hyldgaard}}, \bibinfo
  {author} {\bibfnamefont {A.}~\bibnamefont {Kelkkanen}}, \bibinfo {author}
  {\bibfnamefont {J.}~\bibnamefont {Kleis}}, \bibinfo {author} {\bibfnamefont
  {L.}~\bibnamefont {Kong}}, \bibinfo {author} {\bibfnamefont {S.}~\bibnamefont
  {Li}}, \bibinfo {author} {\bibfnamefont {P.~G.}\ \bibnamefont {Moses}},
  \bibinfo {author} {\bibfnamefont {E.}~\bibnamefont {Murray}}, \bibinfo
  {author} {\bibfnamefont {A.}~\bibnamefont {Puzder}}, \bibinfo {author}
  {\bibfnamefont {H.}~\bibnamefont {Rydberg}}, \bibinfo {author} {\bibfnamefont
  {E.}~\bibnamefont {Schröder}}, \ and\ \bibinfo {author} {\bibfnamefont
  {T.}~\bibnamefont {Thonhauser}},\ }\href
  {http://stacks.iop.org/0953-8984/21/i=8/a=084203} {\bibfield  {journal}
  {\bibinfo  {journal} {J. Phys.: Condens. Matter}\ }\textbf {\bibinfo {volume}
  {21}},\ \bibinfo {pages} {084203} (\bibinfo {year} {2009})}\BibitemShut
  {NoStop}%
\bibitem [{\citenamefont {Rom\'{a}n-P\'{e}rez}\ and\ \citenamefont
  {Soler}(2009)}]{RS09}%
  \BibitemOpen
  \bibfield  {author} {\bibinfo {author} {\bibfnamefont {G.}~\bibnamefont
  {Rom\'{a}n-P\'{e}rez}}\ and\ \bibinfo {author} {\bibfnamefont {J.~M.}\
  \bibnamefont {Soler}},\ }\href {\doibase 10.1103/PhysRevLett.103.096102}
  {\bibfield  {journal} {\bibinfo  {journal} {Phys. Rev. Lett.}\ }\textbf
  {\bibinfo {volume} {103}},\ \bibinfo {pages} {096102} (\bibinfo {year}
  {2009})}\BibitemShut {NoStop}%
\bibitem [{\citenamefont {King-Smith}\ and\ \citenamefont
  {Vanderbilt}(1993)}]{King-Smith1993}%
  \BibitemOpen
  \bibfield  {author} {\bibinfo {author} {\bibfnamefont {R.~D.}\ \bibnamefont
  {King-Smith}}\ and\ \bibinfo {author} {\bibfnamefont {D.}~\bibnamefont
  {Vanderbilt}},\ }\href@noop {} {\bibfield  {journal} {\bibinfo  {journal}
  {Phys. Rev. B}\ }\textbf {\bibinfo {volume} {47}},\ \bibinfo {pages} {1651}
  (\bibinfo {year} {1993})}\BibitemShut {NoStop}%
\bibitem [{\citenamefont {Henkelman}\ \emph {et~al.}(2000)\citenamefont
  {Henkelman}, \citenamefont {Uberuaga},\ and\ \citenamefont
  {Jónsson}}]{Henkelman2000}%
  \BibitemOpen
  \bibfield  {author} {\bibinfo {author} {\bibfnamefont {G.}~\bibnamefont
  {Henkelman}}, \bibinfo {author} {\bibfnamefont {B.~P.}\ \bibnamefont
  {Uberuaga}}, \ and\ \bibinfo {author} {\bibfnamefont {H.}~\bibnamefont
  {Jónsson}},\ }\href {\doibase 10.1063/1.1329672} {\bibfield  {journal}
  {\bibinfo  {journal} {J. Chem. Phys.}\ }\textbf {\bibinfo {volume} {113}},\
  \bibinfo {pages} {9901} (\bibinfo {year} {2000})}\BibitemShut {NoStop}%
\bibitem [{\citenamefont {Olsen}\ \emph {et~al.}(2004)\citenamefont {Olsen},
  \citenamefont {Kroes}, \citenamefont {Henkelman}, \citenamefont
  {Arnaldsson},\ and\ \citenamefont {J\'{o}nsson}}]{Olsen2004}%
  \BibitemOpen
  \bibfield  {author} {\bibinfo {author} {\bibfnamefont {R.~A.}\ \bibnamefont
  {Olsen}}, \bibinfo {author} {\bibfnamefont {G.~J.}\ \bibnamefont {Kroes}},
  \bibinfo {author} {\bibfnamefont {G.}~\bibnamefont {Henkelman}}, \bibinfo
  {author} {\bibfnamefont {A.}~\bibnamefont {Arnaldsson}}, \ and\ \bibinfo
  {author} {\bibfnamefont {H.}~\bibnamefont {J\'{o}nsson}},\ }\href {\doibase
  10.1063/1.1809574} {\bibfield  {journal} {\bibinfo  {journal} {J. Chem.
  Phys.}\ }\textbf {\bibinfo {volume} {121}},\ \bibinfo {pages} {9776}
  (\bibinfo {year} {2004})}\BibitemShut {NoStop}%
\bibitem [{\citenamefont {Gonze}\ and\ \citenamefont {Lee}(1997)}]{Gonze1997}%
  \BibitemOpen
  \bibfield  {author} {\bibinfo {author} {\bibfnamefont {X.}~\bibnamefont
  {Gonze}}\ and\ \bibinfo {author} {\bibfnamefont {C.}~\bibnamefont {Lee}},\
  }\href {\doibase 10.1103/PhysRevB.55.10355} {\bibfield  {journal} {\bibinfo
  {journal} {Phys. Rev. B}\ }\textbf {\bibinfo {volume} {55}},\ \bibinfo
  {pages} {10355} (\bibinfo {year} {1997})}\BibitemShut {NoStop}%
\bibitem [{\citenamefont {Baroni}\ \emph {et~al.}(2001)\citenamefont {Baroni},
  \citenamefont {de~Gironcoli},\ and\ \citenamefont {{Dal Corso}}}]{BGD01}%
  \BibitemOpen
  \bibfield  {author} {\bibinfo {author} {\bibfnamefont {S.}~\bibnamefont
  {Baroni}}, \bibinfo {author} {\bibfnamefont {S.}~\bibnamefont
  {de~Gironcoli}}, \ and\ \bibinfo {author} {\bibfnamefont {A.}~\bibnamefont
  {{Dal Corso}}},\ }\href {\doibase 10.1103/RevModPhys.73.515} {\bibfield
  {journal} {\bibinfo  {journal} {Rev. Mod. Phys.}\ }\textbf {\bibinfo {volume}
  {73}},\ \bibinfo {pages} {515} (\bibinfo {year} {2001})}\BibitemShut
  {NoStop}%
\bibitem [{\citenamefont {Sadhukhan}\ \emph {et~al.}(1999)\citenamefont
  {Sadhukhan}, \citenamefont {Mu\~{n}oz}, \citenamefont {Adamo},\ and\
  \citenamefont {Scuseria}}]{Sadhukhan1999}%
  \BibitemOpen
  \bibfield  {author} {\bibinfo {author} {\bibfnamefont {S.}~\bibnamefont
  {Sadhukhan}}, \bibinfo {author} {\bibfnamefont {D.}~\bibnamefont
  {Mu\~{n}oz}}, \bibinfo {author} {\bibfnamefont {C.}~\bibnamefont {Adamo}}, \
  and\ \bibinfo {author} {\bibfnamefont {G.~E.}\ \bibnamefont {Scuseria}},\
  }\href {\doibase 10.1016/S0009-2614(99)00442-X} {\bibfield  {journal}
  {\bibinfo  {journal} {Chem. Phys. Lett.}\ }\textbf {\bibinfo {volume}
  {306}},\ \bibinfo {pages} {83} (\bibinfo {year} {1999})}\BibitemShut
  {NoStop}%
\bibitem [{\citenamefont {Zhao}\ \emph {et~al.}(2004)\citenamefont {Zhao},
  \citenamefont {Lynch},\ and\ \citenamefont {Truhlar}}]{Zhao2004}%
  \BibitemOpen
  \bibfield  {author} {\bibinfo {author} {\bibfnamefont {Y.}~\bibnamefont
  {Zhao}}, \bibinfo {author} {\bibfnamefont {B.~J.}\ \bibnamefont {Lynch}}, \
  and\ \bibinfo {author} {\bibfnamefont {D.~G.}\ \bibnamefont {Truhlar}},\
  }\href {\doibase 10.1021/jp049908s} {\bibfield  {journal} {\bibinfo
  {journal} {J. Phys. Chem. A}\ }\textbf {\bibinfo {volume} {108}},\ \bibinfo
  {pages} {2715} (\bibinfo {year} {2004})}\BibitemShut {NoStop}%
\end{thebibliography}%

\end{document}